\documentclass[11pt]{article}
\usepackage[left=1in,top=1in,right=1in,bottom=1in]{geometry}
\usepackage{times}
\usepackage{verbatim}
\usepackage{amsmath}
\usepackage{rotating}
\usepackage{algorithm}
\usepackage[noend]{algpseudocode}

\usepackage{silence}
\usepackage{tabularx}
\usepackage{graphicx}
\usepackage{url}
\usepackage{multirow}
\usepackage{subfig}

\usepackage{siunitx}
\usepackage{amssymb}

\allowdisplaybreaks

\begin{document}
\title{Fast Complete Algorithm for Multiplayer Nash Equilibrium}
\author{Sam Ganzfried\\
Ganzfried Research\\
sam@ganzfriedresearch.com
%College of Engineering and Computing \\
%11200 S.W. 8th St. Modesto A. Maidique Campus, ECS 381 \\
%Miami, FL 33199
}

\date{\vspace{-5ex}}

\maketitle

\begin{abstract}
We describe a new complete algorithm for computing Nash equilibrium in multiplayer general-sum games, based on a quadratically-constrained feasibility program formulation. We demonstrate that the algorithm runs significantly faster than the prior fastest complete algorithm on several game classes previously studied and that its runtimes even outperform the best incomplete algorithms.
\end{abstract}

\section{Introduction}
\label{se:intro}
Nash equilibrium is the central solution concept in game theory. While a Nash equilibrium can be computed in polynomial time for two-player zero-sum games, it is PPAD-hard for two-player general-sum and multiplayer games and widely believed that no efficient algorithms exist~\cite{Chen05:Nash,Chen06:Settling,Daskalakis09:Complexity}. Furthermore, even if we were able to compute an equilibrium for these game classes, it would have no performance guarantee. In a two-player zero-sum game, every Nash equilibrium guarantees at least the value of the game in expectation in the worst case. Therefore, if players alternate roles, a Nash equilibrium would guarantee a win or tie in expectation regardless of the strategy used by the opponent. However, for non-zero-sum and multiplayer games, an equilibrium would have no performance guarantee. There can be multiple equilibria with different values, and if the opponents play strategies from a different equilibrium than ours then the resulting strategies may not be in equilibrium.

Despite these computational and conceptual challenges, we must still create agents with strong strategies for these settings, and Nash equilibrium is a compelling starting point. It was shown that an exact Nash equilibrium strategy defeated a variety of agents submitted for a class project in 3-player Kuhn poker~\cite{Ganzfried18:Successful}. Recently an agent was created for 6-player no-limit Texas hold 'em that defeated strong human players by attempting to approximate Nash equilibrium strategies~\cite{Brown19:Superhuman}. The core equilibrium-finding technique used by this agent was based on the counterfactual regret minimization algorithm, an iterative self-play procedure~\cite{Zinkevich07:Regret}. It has been demonstrated that counterfactual regret minimization does in fact converge to an $\epsilon$-Nash equilibrium (a strategy profile in which no player can gain more than $\epsilon$ by deviating) for small $\epsilon$ in three-player Kuhn poker, while it does not converge to equilibrium in the larger game of three-player Leduc hold 'em~\cite{Abou10:Using}. These results show that Nash equilibrium strategies (or their approximations) can be successful in practice despite the fact that they do not have a performance guarantee.

Several algorithms have been developed for computing Nash equilibrium in multiplayer games; however, many of them are incomplete, slow, and/or produce solutions with poor approximation quality (i.e., high $\epsilon$). An algorithm is \emph{complete} if it always finds a solution when one exists (at least one Nash equilibrium is guaranteed to exist in all finite games~\cite{Nash50:Non-cooperative}). We present a new algorithm that is complete and runs significantly faster than prior complete algorithms, and even runs faster than the best incomplete algorithms. Our algorithm is based on a novel quadratically-constrained mixed-integer program formulation that utilizes a technique from the newest Gurobi release~\cite{Gurobi19:Gurobi}. We run experiments on uniform random games with a variety of players and strategy sizes, as well as several games produced from the GAMUT generator~\cite{Nudelman04:Run}. We compare our algorithm against the best prior algorithms, which include several complete methods as well as faster incomplete methods available on the GAMBIT software suite~\cite{McKelvey04:Gambit}.

\section{Notation}
\label{se:notation}
A strategic-form game consists of a finite set of players $N = \{1,\ldots,n\}$, a finite set of pure strategies $S_i$ for each player $i \in N$, and a real-valued utility for each player for each strategy vector (aka \emph{strategy profile}), $u_i : \times_i S_i \rightarrow \mathbb{R}$. We will assume that the sets $S_i$ are disjoint, and for simplicity assume that all $S_i$ have the same cardinality. For $s_j \in S_i$ define the \emph{player} function to be $P(s_j) = i$ (which is well-defined under the assumption that the $S_i$ are disjoint). Suppose that $s_{j_k} \in S_{i_k}$ for $k = 1 \ldots n,$ and suppose that the $i_k \in N$ are all distinct. Then for $w \in N$ define 
$\hat{u}_w(s_{j_1},\ldots,s_{j_n}) = u_w(s_{m_1},\ldots,s_{m_n}),$
where $m_q$ equals the $j_k$ such that $P(j_k) = q$ (and therefore that $s_{m_q} \in S_q$). That is, in the event that the $s_{j_k}$ are not in order of increasing value of the player $P(s_{j_k})$, the $\hat{u}$ function will compute the utility assuming that the vector of strategies is listed in the order of increasing players so that $u$ can be properly applied. For example, suppose that $s_1 \in S_1, s_2 \in S_2, s_3 \in S_3.$ Then $\hat{u}_w(s_2,s_3,s_1) = u_w(s_1,s_2,s_3),$ for $w \in N$. This notation will be useful in order to provide more concise representations of our optimization formulations.

A \emph{mixed strategy} $\sigma_i$ for player $i$ is a probability distribution over pure strategies, where $\sigma_i(s_{i'})$ is the probability that player $i$ plays pure strategy $s_{i'} \in S_i$ under $\sigma_i$. Let $\Sigma_i$ denote the full set of mixed strategies for player $i$. A strategy profile $\sigma^* = (\sigma^*_1,\ldots,\sigma^*_n)$ is a \emph{Nash equilibrium} if $u_i(\sigma^*_i,\sigma^*_{-i}) \geq u_i(\sigma_i, \sigma^*_{-i})$ for all $\sigma_i \in \Sigma_i$ for all $i \in N$, where $\sigma^*_{-i}$ denotes the vector of the components of strategy $\sigma^*$ for all players excluding player i. For a given candidate strategy profile $\sigma^*$, define $\epsilon = \epsilon(\sigma^*) = \max_i \max_{\sigma_i \in \Sigma_i} \left[ u_i(\sigma_i,\sigma^*_{-i}) - u_i(\sigma^*_i, \sigma^*_{-i}) \right]$.

\section{Algorithm}
\label{se:alg}
We first describe a linear mixed-integer feasibility program formulation for computing Nash equilibrium in two-player general-sum games~\cite{Sandholm05:Mixed}. That work presented four different formulations each using a different objective function and set of constraints, and demonstrated that the first one significantly outperformed the other three. The first formulation was a feasibility program with no objective function, in which the set of Nash equilibria correspond exactly to feasible solutions. We use this formulation as a starting point for our new multiplayer formulations.

\subsection{Linear mixed-integer feasibility formulation for two-player Nash equilibrium}
\label{se:2-player}
We quote from the original description of the program formulation for two-player Nash equilibrium, and present the formulation below:

\begin{quote}
In our first formulation, the feasible solutions are exactly the equilibria of the game.  For every pure strategy $s_i$, there is  binary variable $b_{s_i}$.  If this variable is set to 1, the probability placed on the strategy must be 0. If it is set to 0, the strategy is allowed to be in the support, but the regret of the strategy must be 0. The formulation has the following variables other than the $b_{s_i}$.  For each player, there is a variable $u_i$ indicating the highest possible expected utility that that player  can obtain given the other player's mixed strategy. For every pure strategy $s_i$, there is a variable $p_{s_i}$ indicating the probability placed on that strategy, a variable $u_{s_i}$ indicating the expected utility of playing that strategy (given the other player's mixed  strategy), and a variable $r_{s_i}$ indicating the regret of playing $s_i$.  The constant $U_i$ indicates the maximum difference between two utilities in the game for player $i$: $U_i = \max_{s^h_i, s^l_i \in S_i, s^h_{1-i},s^l_{1-i} \in S_{1-i}} \left[ u_i(s^h_i, s^h_{1-i}) - u_i(s^l_i, s^l_{1-i}) \right].$ The formulation follows below~\cite{Sandholm05:Mixed}.
\end{quote}

Find $p_{s_i},u_i,u_{s_i},r_{s_i},b_{s_i}$ subject to:
\begin{align}
&\sum_{s_i \in S_i} p_{s_i} = 1 \mbox{ for all } i \\
&u_{s_i} = \sum_{s_{1-i} \in S_{1-i}} p_{s_{1-i}}u_i(s_i,s_{1-i}) \mbox{ for all } i, s_i \in S_i \\
&u_i \geq u_{s_i} \mbox{ for all } i, s_i \in S_i \label{eq:2p-u} \\
&r_{s_i} = u_i - u_{s_i} \mbox{ for all } i, s_i \in S_i \label{eq:2p-r1}\\
&p_{s_i} \leq 1 - b_{s_i} \mbox{ for all } i, s_i \in S_i \label{eq:2p-p}\\
&r_{s_i} \leq U_i b_{s_i} \mbox{ for all } i, s_i \in S_i \label{eq:2p-r2}\\
&p_{s_i} \geq 0 \mbox{ for all } i, s_i \in S_i \\
&u_i \geq 0 \mbox{ for all } i \\
&u_{s_i} \geq 0 \mbox{ for all } i, s_i \in S_i \\
&r_{s_i} \geq 0 \mbox{ for all } i, s_i \in S_i \label{eq:2p-rbound}\\
&b_{s_i} \mbox{ binary in } \{0,1\} \mbox{ for all } i, s_i \in S_i 
\end{align}

\begin{quote}
The first four constraints ensure that the $p_{s_i}$ values constitute a valid probability distribution and define the regret of a strategy. Constraint~\ref{eq:2p-p} ensures that $b_{s_i}$ can be set to 1 only when no probability is placed on $s_i$. On the other hand, Constraint~\ref{eq:2p-r2} ensures that the regret of a strategy equals 0, unless $b_{s_i} = 1$,  in which  case  the constraint is vacuous because the regret can never exceed $U_i$. (Technically, Constraint~\ref{eq:2p-u} is redundant as it follows from Constraints~\ref{eq:2p-r1} and~\ref{eq:2p-rbound}.)~\cite{Sandholm05:Mixed}
\end{quote}

For clarity, we will rewrite the system with the redundant Constraint~\ref{eq:2p-u} removed, as our extensions will be based on this formulation.

Find $p_{s_i},u_i,u_{s_i},r_{s_i},b_{s_i}$ subject to:
\begin{align*}
&\sum_{s_i \in S_i} p_{s_i} = 1 \mbox{ for all } i\\
&u_{s_i} = \sum_{s_{1-i} \in S_{1-i}} p_{s_{1-i}}u_i(s_i,s_{1-i}) \mbox{ for all } i, s_i \in S_i\\
&r_{s_i} = u_i - u_{s_i} \mbox{ for all } i, s_i \in S_i\\
&p_{s_i} \leq 1 - b_{s_i} \mbox{ for all } i, s_i \in S_i \\
&r_{s_i} \leq U_i b_{s_i} \mbox{ for all } i, s_i \in S_i \\
&p_{s_i} \geq 0 \mbox{ for all } i, s_i \in S_i \\
&u_i \geq 0 \mbox{ for all } i\\
&u_{s_i} \geq 0 \mbox{ for all } i, s_i \in S_i \\
&r_{s_i} \geq 0 \mbox{ for all } i, s_i \in S_i \\
&b_{s_i} \mbox{ binary in } \{0,1\} \mbox{ for all } i, s_i \in S_i
\end{align*}

\subsection{New formulation for three-player Nash equilibrium}
\label{se:3-player}
We now describe an extension of the previous two-player formulation to three players. To do this, we introduce new variables, $p_{s_i,s_j}$, 
which denote the product of the variables $p_{s_i}$ and $p_{s_j}$. Note that these new product constraints are now quadratic (while all other constraints remain linear). 

Find $p_{s_i},u_i,u_{s_i},r_{s_i},b_{s_i},p_{s_i,s_j}$ subject to:
\begin{align*}
&\sum_{s_i \in S_i} p_{s_i} = 1 \mbox{ for all } i \\
&u_{s_i} = \sum_{s_j \in S_2} \sum_{s_k \in S_3} p_{s_j,s_k} u_1(s_i,s_j,s_k) \mbox{ for all } s_i \in S_1\\
&u_{s_j} = \sum_{s_i \in S_1} \sum_{s_k \in S_3} p_{s_i,s_k} u_2(s_i,s_j,s_k) \mbox{ for all } s_j \in S_2\\
&u_{s_k} = \sum_{s_i \in S_1} \sum_{s_j \in S_2} p_{s_i,s_j} u_3(s_i,s_j,s_k) \mbox{ for all } s_k \in S_3\\
&p_{s_i,s_j} = p_{s_i} \cdot p_{s_j}  \mbox{ for all } s_i \in S_1, s_j \in S_2\\
&p_{s_i,s_j} = p_{s_i} \cdot p_{s_j}  \mbox{ for all } s_i \in S_1, s_j \in S_3\\
&p_{s_i,s_j} = p_{s_i} \cdot p_{s_j}  \mbox{ for all } s_i \in S_2, s_j \in S_3\\
&r_{s_i} = u_i - u_{s_i} \mbox{ for all } i, s_i \in S_i\\
&p_{s_i} \leq 1 - b_{s_i} \mbox{ for all } i, s_i \in S_i \\
&r_{s_i} \leq U_i b_{s_i} \mbox{ for all } i, s_i \in S_i \\
&p_{s_i} \geq 0 \mbox{ for all } i, s_i \in S_i \\
&u_i \geq 0 \mbox{ for all } i\\
&u_{s_i} \geq 0 \mbox{ for all } i, s_i \in S_i \\
&r_{s_i} \geq 0 \mbox{ for all } i, s_i \in S_i \\
&b_{s_i} \mbox{ binary in } \{0,1\} \mbox{ for all } i, s_i \in S_i
\end{align*}

We can simplify the presentation by condensing the constraints for $u_{s_i}$ and for the product variables $p_{s_i,s_j}$, using the notation for $\hat{u}$ defined in Section~\ref{se:notation}.

Find $p_{s_i},u_i,u_{s_i},r_{s_i},b_{s_i},p_{s_i,s_j}$ subject to:
\begin{align*}
&\sum_{s_i \in S_i} p_{s_i} = 1 \mbox{ for all } i \\
&u_{s_i} = \sum_{s_j \in S_J} \sum_{s_k \in S_K} p_{s_j,s_k} \hat{u}_{P(s_i)}(s_i,s_j,s_k) \mbox{ for all } I, J \neq I, K \neq I, J < K, s_i \in S_I\\
&p_{s_i,s_j} = p_{s_i} \cdot p_{s_j} \mbox{ for all } I, J \in N, I < J, s_i \in S_I, s_j \in S_J\\
&r_{s_i} = u_i - u_{s_i} \mbox{ for all } i, s_i \in S_i\\
&p_{s_i} \leq 1 - b_{s_i} \mbox{ for all } i, s_i \in S_i \\
&r_{s_i} \leq U_i b_{s_i} \mbox{ for all } i, s_i \in S_i \\
&p_{s_i} \geq 0 \mbox{ for all } i, s_i \in S_i \\
&u_i \geq 0 \mbox{ for all } i\\
&u_{s_i} \geq 0 \mbox{ for all } i, s_i \in S_i \\
&r_{s_i} \geq 0 \mbox{ for all } i, s_i \in S_i \\
&b_{s_i} \mbox{ binary in } \{0,1\} \mbox{ for all } i, s_i \in S_i
\end{align*}

\subsection{New formulation for four-player Nash equilibrium}
\label{se:4-player}
We further extend our 3-player formulation to 4 players by introducing new variables $p_{s_i,s_j,s_k}.$ We still retain the $p_{s_i,s_j}$ variables as before, and include additional constraints of the form $p_{s_i,s_j,s_k} = p_{s_i} \cdot p_{s_j,s_k}.$ Thus, despite the expected utilities being cubic in the original variables $p_{s_i}$, we are able to obtain a formulation that only has linear and quadratic constraints.

Find $p_{s_i},u_i,u_{s_i},r_{s_i},b_{s_i},p_{s_i,s_j},p_{s_i,s_j,s_k}$ subject to:
\begin{align*}
&\sum_{s_i \in S_i} p_{s_i} = 1 \mbox{ for all } i \\
&u_{s_i} = \sum_{s_j \in S_2, s_k \in S_3, s_m \in S_4} p_{s_j,s_k,s_m} u_1(s_i,s_j,s_k,s_m) \mbox{ for all } s_i \in S_1\\
&u_{s_j} = \sum_{s_i \in S_1, s_k \in S_3, s_m \in S_4} p_{s_i,s_k,s_m} u_2(s_i,s_j,s_k,s_m) \mbox{ for all } s_j \in S_2\\
&u_{s_k} = \sum_{s_i \in S_1, s_j \in S_2, s_m \in S_4} p_{s_i,s_j,s_m} u_3(s_i,s_j,s_k,s_m) \mbox{ for all } s_k \in S_3\\
&u_{s_m} = \sum_{s_i \in S_1, s_j \in S_2, s_k \in S_3} p_{s_i,s_j,s_k} u_4(s_i,s_j,s_k,s_m) \mbox{ for all } s_m \in S_4\\
&p_{s_i,s_j} = p_{s_i} \cdot p_{s_j}  \mbox{ for all } s_i \in S_1, s_j \in S_2\\
&p_{s_i,s_j} = p_{s_i} \cdot p_{s_j}  \mbox{ for all } s_i \in S_1, s_j \in S_3\\
&p_{s_i,s_j} = p_{s_i} \cdot p_{s_j}  \mbox{ for all } s_i \in S_1, s_j \in S_4\\
&p_{s_i,s_j} = p_{s_i} \cdot p_{s_j}  \mbox{ for all } s_i \in S_2, s_j \in S_3\\
&p_{s_i,s_j} = p_{s_i} \cdot p_{s_j}  \mbox{ for all } s_i \in S_2, s_j \in S_4\\
&p_{s_i,s_j} = p_{s_i} \cdot p_{s_j}  \mbox{ for all } s_i \in S_3, s_j \in S_4\\
&p_{s_i,s_j,s_k} = p_{s_i} \cdot p_{s_j,s_k}  \mbox{ for all } s_i \in S_1, s_j \in S_2, s_k \in S_3\\
&p_{s_i,s_j,s_k} = p_{s_i} \cdot p_{s_j,s_k}  \mbox{ for all } s_i \in S_1, s_j \in S_2, s_k \in S_4\\
&p_{s_i,s_j,s_k} = p_{s_i} \cdot p_{s_j,s_k}  \mbox{ for all } s_i \in S_1, s_j \in S_3, s_k \in S_4\\
&p_{s_i,s_j,s_k} = p_{s_i} \cdot p_{s_j,s_k}  \mbox{ for all } s_i \in S_2, s_j \in S_3, s_k \in S_4\\
&r_{s_i} = u_i - u_{s_i} \mbox{ for all } i, s_i \in S_i\\
&p_{s_i} \leq 1 - b_{s_i} \mbox{ for all } i, s_i \in S_i \\
&r_{s_i} \leq U_i b_{s_i} \mbox{ for all } i, s_i \in S_i \\
&p_{s_i} \geq 0 \mbox{ for all } i, s_i \in S_i \\
&u_i \geq 0 \mbox{ for all } i\\
&u_{s_i} \geq 0 \mbox{ for all } i, s_i \in S_i \\
&r_{s_i} \geq 0 \mbox{ for all } i, s_i \in S_i \\
&b_{s_i} \mbox{ binary in } \{0,1\} \mbox{ for all } i, s_i \in S_i
\end{align*}

As for the 3-player version we can simplify the presentation by condensing constraints and utilizing $\hat{u}$.

Find $p_{s_i},u_i,u_{s_i},r_{s_i},b_{s_i},p_{s_i,s_j},p_{s_i,s_j,s_k}$ subject to:
\begin{align*}
&\sum_{s_i \in S_i} p_{s_i} = 1 \mbox{ for all } i \\
&u_{s_i} = \sum_{s_j \in S_J} \sum_{s_k \in S_K} \sum_{s_m \in S_M} p_{s_j,s_k,s_m} \hat{u}_{P(s_i)}(s_i,s_j,s_k,s_m) \mbox{ for all } I, J \neq I, K \neq I, M \neq I, J < K < M, s_i \in S_I\\
&p_{s_i,s_j} = p_{s_i} \cdot p_{s_j}  \mbox{ for all } I, J \in N, I < J, s_i \in S_I, s_j \in S_J\\
&p_{s_i,s_j,s_k} = p_{s_i} \cdot p_{s_j,s_k}  \mbox{ for all } I, J, K \in N, I < J < K, s_i \in S_I, s_j \in S_J, s_k \in S_K\\
&r_{s_i} = u_i - u_{s_i} \mbox{ for all } i, s_i \in S_i\\
&p_{s_i} \leq 1 - b_{s_i} \mbox{ for all } i, s_i \in S_i \\
&r_{s_i} \leq U_i b_{s_i} \mbox{ for all } i, s_i \in S_i \\
&p_{s_i} \geq 0 \mbox{ for all } i, s_i \in S_i \\
&u_i \geq 0 \mbox{ for all } i\\
&u_{s_i} \geq 0 \mbox{ for all } i, s_i \in S_i \\
&r_{s_i} \geq 0 \mbox{ for all } i, s_i \in S_i \\
&b_{s_i} \mbox{ binary in } \{0,1\} \mbox{ for all } i, s_i \in S_i
\end{align*}

\subsection{Five-player Nash equilibrium}
\label{se:5-player}
We can create a similar extension for 5 players that again only uses linear and quadratic constraints.

Find $p_{s_i},u_i,u_{s_i},r_{s_i},b_{s_i},p_{s_i,s_j},p_{s_i,s_j,s_k},p_{s_i,s_j,s_k,s_m}$ subject to:
\small
\begin{align*}
&\sum_{s_i \in S_i} p_{s_i} = 1 \mbox{ for all } i \\
&u_{s_i} = \sum_{s_j \in S_J} \sum_{s_k \in S_K} \sum_{s_m \in S_M} \sum_{s_o \in S_O} p_{s_j,s_k,s_m,s_o} \hat{u}_{P(s_i)}(s_i,s_j,s_k,s_m,s_o) \forall I, \{J,K,M,O\} \neq I, J < K < M < O, s_i \in S_I\\
&p_{s_i,s_j} = p_{s_i} \cdot p_{s_j}  \mbox{ for all } I, J \in N, I < J, s_i \in S_I, s_j \in S_J\\
&p_{s_i,s_j,s_k} = p_{s_i} \cdot p_{s_j,s_k}  \mbox{ for all } I, J, K \in N, I < J < K, s_i \in S_I, s_j \in S_J, s_k \in S_K\\
&p_{s_i,s_j,s_k,s_m} = p_{s_i} \cdot p_{s_j,s_k,s_m}  \mbox{ for all } I, J, K, M \in N, I < J < K < M, s_i \in S_I, s_j \in S_J, s_k \in S_K, s_m \in S_M\\
&r_{s_i} = u_i - u_{s_i} \mbox{ for all } i, s_i \in S_i\\
&p_{s_i} \leq 1 - b_{s_i} \mbox{ for all } i, s_i \in S_i \\
&r_{s_i} \leq U_i b_{s_i} \mbox{ for all } i, s_i \in S_i \\
&p_{s_i} \geq 0 \mbox{ for all } i, s_i \in S_i \\
&u_i \geq 0 \mbox{ for all } i\\
&u_{s_i} \geq 0 \mbox{ for all } i, s_i \in S_i \\
&r_{s_i} \geq 0 \mbox{ for all } i, s_i \in S_i \\
&b_{s_i} \mbox{ binary in } \{0,1\} \mbox{ for all } i, s_i \in S_i
\end{align*}
\normalsize

\subsection{New formulation for n-player Nash equilibrium}
\label{se:k-player}
One can easily see how our formulation can be generalized to one for $n$ players that has only linear and quadratic constraints. There will be $m^k\binom{n}{k}$ of the $p_{s_{i_1},\ldots,s_{i_k}}$ terms of length $k$ for each $1 \leq k \leq n-1,$ where $m = |S_i|$ is the number of pure strategies for each player. So the total number of the $p$ terms will be $\sum_{k = 1} ^ {n-1} m^k \binom{n}{k}.$ From the binomial theorem, we know that $(1+m)^n = \sum_{k = 0} ^ n m^k \binom{n}{k}.$
So the total number of the $p$ terms is
\begin{eqnarray*}
&& (1+m)^n - m^0 \binom{n}{0} - m^n \binom{n}{n} \\
&= &(1+m)^n - 1 - m^n \\ 
&< &(1+m)^n\\
\end{eqnarray*}
Note that while this is exponential in the number of players, the size of the game representation is $n \cdot m^n$, since we must specify a payoff for each player for each of $m^n$ pure strategy profiles, which is also exponential in the number of players.

\subsection{Computation of Nash equilibrium from new quadratic program formulation}
\label{se:MIQCP}
While we have been able to formulate the problem of computing a Nash equilibrium for $n \geq 3$ players as a quadratically-constrained program (QCP), unfortunately the constraint matrix is not positive semidefinite making the overall program non-convex and more challenging to solve. The best commercial solvers could previously solve convex QCPs but not non-convex QCPs, and the best approach was to approximate products of variables by using piecewise linear approximations~\cite{Bisschop06:AIMMS}; however, this approach introduces a large number of new variables and constraints, leading to large run times, as well as an added layer of approximation error. Recently Gurobi has released an approach that is able to solve non-convex programs with quadratic objective and constraints~\cite{Gurobi19:Gurobi}. The solver allows for both continuous and integral variables, and so can handle mixed-integer quadratically-constrained programs (MIQCPs), which is what we are interested in. The new method addresses non-convex bilinear constraints using an analogue of the simplex algorithm with McCormick envelopes for constructing relaxations with new approaches for cutting planes and spatial branching.

\section{Experiments}
\label{se:exp}
For our first set of experiments, we generated games with payoffs uniformly random in [0,1] for a variety of number of players $n$ and number of pure strategies $m$. We used the same parameter values as those used for previous experiments for complete algorithms~\cite{Berg17:Exclusion}. For each set of parameter values $(n,m)$, we generated 1,000 random games as the prior work had done (with the exception of the largest game $n=5,m=3$ for which we generated 100 games). We set a time limit of 900 seconds for the random game experiments as the prior work had done. 

For all experiments with our algorithm we used Gurobi's non-convex MIQCP solver, with feasibility tolerance parameter set to 0.0001. For the GAMUT experiments we set the NumericFocus parameter to 2. We used an Intel Core i7-8550U at 1.80 GHz with 16 GB of RAM under 64-bit Windows 10 (8 threads). Prior experiments had been done with similar hardware: Intel Core i7-6500U at 2.50 GHz with 16 GB of RAM under 64-bit Windows 7~\cite{Berg17:Exclusion}.

The results from experiments with our MIQCP algorithm on random games are shown in Table~\ref{ta:results-random}. For all games other than the largest class ($n=5,m=3$) the algorithm had very fast runtimes (in most cases averaging a fraction of a second), with zero runs over the time limit. For the largest class the algorithm hit the time limit in 58\% of instances. Analogous results for the best prior complete algorithms are shown in Table~\ref{ta:results-random-old}. Other than for the $n=5,m=3$ games, our algorithm outperformed both other algorithms by orders of magnitude in runtime. 

\begin{table}[!ht]
\centering
\begin{tabular}{|*{5}{c|}} \hline
$n$ &$m$ &Avg. time(s) &Median time(s) &OverTime\% \\ \hline
3 &2 &0.00707 &0.0 &0 \\ \hline
3 &3 &0.02342 &0.02901 &0 \\ \hline
3 &5 &0.85763 &0.26544 &0 \\ \hline
4 &2 &0.02598 &0.03124 &0 \\ \hline
4 &3 &1.35334 &0.40505 &0 \\ \hline
5 &2 &0.11873 &0.09373 &0 \\ \hline
5 &3 &607.68524 &900.0 &58 \\ \hline
\end{tabular}
\caption{Results of new MIQCP algorithm for random games.}
\label{ta:results-random}
\end{table}

\begin{table}[!ht]
\centering
\begin{tabular}{|*{8}{c|}} \hline
\multicolumn{2}{|c|}{} &\multicolumn{3}{|c|}{Exclusion Method} &\multicolumn{3}{|c|}{$k$-Uniform Search} \\ \hline 
$n$ &$m$ &Avg. time(s) &Median time(s) &OverTime\% &$k$ &Avg. time(s) &OverTime\% \\ \hline
3 &2 &0.04 &0.02 &0 &2/180 &49 &1 \\ \hline
3 &3 &26 &1.2 &1 &3/18 &191 &29 \\ \hline
3 &5 &900 &900 &100 &5 &94 &33 \\ \hline
4 &2 &99 &0.48 &8 &2/40 &23 &15 \\ \hline
4 &3 &352 &87 &30 &3/8 &85 &33 \\ \hline
5 &2 &125 &2.7 &10 &2/8 &1.0 &30 \\ \hline
5 &3 &520 &589 &46 &3 &7.9 &36 \\ \hline
\end{tabular}
\caption{Results of prior complete algorithms for random games~\cite{Berg17:Exclusion}.}
\label{ta:results-random-old}
\end{table}

The Exclusion Method is a complete tree-search-based method that has the best upper bound with respect to the number of players $n$~\cite{Berg17:Exclusion}. The algorithm divides the search space into smaller regions and examines whether an equilibrium can exist in the region. The $k$-Uniform Search algorithm is based on an improvement to a prior exhaustive complete method~\cite{Babichenko14:Simple} where a search is performed over the space of $k$-uniform strategies for incrementally increasing $k$. (A \emph{$k$-uniform strategy} is a strategy where all probabilities are integer multiples of $\frac{1}{k}$.) This approach was used as a benchmark in prior work~\cite{Berg17:Exclusion}. Note that these algorithms do not involve the use of a commercial solver such as Gurobi.

We next experimented on several games produced from the GAMUT generator~\cite{Nudelman04:Run}. We used the same games and parameter settings as used in prior work~\cite{Berg17:Exclusion}. In particular, we used the variants with 3 players and 3 actions per player. For the congestion game class we used 2 for the number of facilities parameter, and for the covariant game we used $r = -0.5$. All other parameters were generated randomly (as the prior experiments had done). We generated 1,000 games from each class using these distributions. We did not use any time limit for these experiments.

Results for our new MIQCP algorithm over the GAMUT games are shown in Table~\ref{ta:results-gamut}. We normalized all payoffs to be in [0,1] (by subtracting the smallest payoff from all the payoffs and then dividing all payoffs by the difference between the max and min payoff, or just dividing by the max payoff if the min is nonnegative). Note that linear transformations of the payoffs exactly preserve Nash equilibria, so this normalization would have no effect on the solutions. For some classes several games generated had NaN payoff values, and we ignored these games (we report the number of valid games). We can see that our algorithm ran very quickly for all classes and correctly solved all instances. %We note that there was one Random graphical game instance where our algorithm incorrectly output that the program was infeasible due to numerical instability. Despite the completeness of the algorithm in theory, there can always be certain specific games constructed that can lead to numerical issues for optimization solvers that deal with floating-point arithmetic and feasibility tolerances. However, our algorithm correctly solved all instances with this one exception. 

\begin{table}[!ht]
\centering
\begin{tabular}{|*{4}{c|}} \hline
Game class &\# valid games &Avg. time(s) &\# NotSolved \\ \hline
Bertrand oligopoly &970 &0.00106 &0 \\ \hline
Bidirectional LEG &1000 &0.00508 &0\\ \hline
Collaboration &1000 &0.00962 &0\\ \hline
Congestion &1000 &0.00492 &0\\ \hline
Covariant &1000 &0.02984 &0\\ \hline
Polymatrix &997 &0.00803 &0\\ \hline
Random graphical &1000 &0.01615 &0 \\ \hline
Random LEG &1000 &0.00475 &0 \\ \hline
Uniform LEG &1000 &0.00468 &0 \\ \hline
\end{tabular}
\caption{Results of new MIQCP algorithm for GAMUT games.}
\label{ta:results-gamut}
\end{table}

Analogous results for the prior best complete algorithms for these same game classes are in Table~\ref{ta:results-gamut-old}. We can see again that our algorithm typically runs orders of magnitude faster than the others. Note that for these results the NotSolved\% column refers to the percentage of runs where the $\epsilon$ of the computed strategies exceeded 0.001 (this was the criterion from prior work~\cite{Berg17:Exclusion}). 

\begin{table}[!ht]
\centering
\begin{tabular}{|*{8}{c|}} \hline
\multicolumn{1}{|c|}{} &\multicolumn{2}{|c|}{Exclusion Method} &\multicolumn{2}{|c|}{$k$-Uniform Search} \\ \hline 
Game class &Avg. time(s) &NotSolved\% &Avg. time(s) &NotSolved\% \\ \hline
Bertrand oligopoly &13.7 &0 &0.01 &0 \\ \hline
Bidirectional LEG &159 &0 &0.013 &0\\ \hline
Collaboration &2.8 &0 &0.0009 &0\\ \hline
Congestion &29 &0 &0.027 &0\\ \hline
Covariant &95 &0 &80 &16\\ \hline
Polymatrix &172 &0 &27.2 &7\\ \hline
Random graphical &35000 &0 &0.05 &0 \\ \hline
Random LEG &880 &0 &0.02 &0 \\ \hline
Uniform LEG &793 &0 &0.02 &0 \\ \hline
\end{tabular}
\caption{Results of prior complete algorithms for GAMUT games~\cite{Berg17:Exclusion}.}
\label{ta:results-gamut-old}
\end{table}

Table~\ref{ta:results-gambit} shows results for these same game classes using the best algorithms from the GAMBIT software suite~\cite{McKelvey04:Gambit}. The numbers are the average computation times in seconds and the parentheses show the percentage of instances that were not solved (code got stuck, empty output, or accuracy not within the given $\epsilon = 0.001$). All of these methods are incomplete, and in many cases the NotSolved\% was quite large. The runtimes of our algorithm are still about one order of magnitude better than these methods, while also correctly solving all instances.

\begin{table}[!ht]
\centering
\begin{tabular}{|*{7}{c|}} \hline
Game class &gnm &ipa &enumpoly &simpdiv &liap &logit \\ \hline
Bertrand oligopoly &0.05 (30) &0.05 (75) &0.04 (50) &0.05 &0.24 (99) &0.06 \\ \hline
Bidirectional LEG &0.09 (0.3) &0.05 (58) &0.84 (1) &0.06 (0.1) &0.24 (99) &0.06 (0.1)\\ \hline
Collaboration &0.24 (0.1) &0.04 &3.3 (50) &0.05 &0.34 (99) &0.06 (0.3)\\ \hline
Congestion &0.05 (0.2) &0.05 (85) &0.05 (0.6) &0.05 (0.1) &0.21 (100) &0.05\\ \hline
Covariant &0.13 (3) &0.05 (94) &36 &0.67 (2.8) &0.31 (100) &0.05 (1)\\ \hline
Polymatrix &0.06 (1) &0.04 (79) &0.04 (50) &0.07 (0.3) &0.3 (92) &0.05 (0.4)\\ \hline
Random graphical &0.08 (3) &0.04 (96) &6.3 (6) &0.17 (3) &0.31 (99) &0.06 (0.3) \\ \hline
Random LEG &0.05 (1) &0.04 (59) &8.1 (2) &0.05 (0.6) &0.24 (99) &0.06 \\ \hline
Uniform LEG &0.07 (0.4) &0.05 (55) &0.04 (17) &0.05 &0.23 (99) &0.06 \\ \hline
\end{tabular}
\caption{Computation times in seconds for GAMBIT algorithms and \% of instances not solved~\cite{Berg17:Exclusion}.}
\label{ta:results-gambit}
\end{table}

The algorithms are the homotopy method~\cite{Govindan03:Global} (gnm), its modification using iterated polymatrix approximation~\cite{Govindan04:Computing} (ipa), an algorithm based on solving a polynomial system of equations~\cite{Porter08:Simple} (enumpoly), the simplicial subdivision method~\cite{Van87:Simplicial} (simpdiv), a function minimization approach (liap), and the quantal response method~\cite{McKelvey95:Quantal,Turocy05:Dynamic} (logit).\footnote{These experiments were performed using GAMBIT version 15.0 (except 16.0 for simpdiv as it had a bug in 15.0; only simpdiv changed from 15.0 to 16.0, so only that algorithm was rerun with version 16.0)~\cite{Berg17:Exclusion}.}

Our final comparison is with two recently popular algorithms, counterfactual regret minimization~\cite{Zinkevich07:Regret} and fictitious play~\cite{Brown51:Iterative,Robinson51:Iterative}. These are iterative self-play procedures that have been proven to converge to Nash equilibrium in two-player zero-sum games, but not for more than two players. However, they can both be run for more than two players, and have been demonstrated to obtain strong empirical performance in certain large extensive-form imperfect-information games.  
For example, an agent that utilized counterfactual regret minimization (CFR) recently defeated strong humans in 6-player no limit Texas hold 'em~\cite{Brown19:Superhuman}. Both CFR and fictitious play (FP) can be extremely effective at quickly approximating Nash equilibrium strategies, particularly in large games. However, they can also lead to strategies with extremely high $\epsilon$, even for very small games. So if the goal is to compute an exact Nash equilibrium in multiplayer games, CFR and FP are ineffective. Table~\ref{ta:results-fpcfr} shows recent results of CFR and FP for games with uniform random payoffs in [0,1]~\cite{Ganzfried20:Fictitious}. We can see that in several cases the average values of $\epsilon$ are quite large, and in all cases it exceeds the previously designated benchmark value of $0.001$~\cite{Berg17:Exclusion}.    

\begin{table*}[!ht]
\centering
%\footnotesize
%\small
\begin{tabular}{|*{6}{c|}} \hline
$n$ &$m$ &\# games &\# algorithm iterations &Avg. CFR $\epsilon$ &Avg. FP $\epsilon$\\ \hline
3 &3 &100,000 &10,000 &0.00768 &0.00749 \\ \hline
3 &5 &100,000 &10,000 &0.02312 &0.02244 \\ \hline
3 &10 &10,000 &10,000 &0.05963 &0.05574 \\ \hline
4 &3 &100,000 &10,000 &0.01951 &0.01950 \\ \hline
4 &5 &10,000 &10,000 &0.05121 &0.04635 \\ \hline
4 &10 &10,000 &10,000 &0.08315 &0.06661 \\ \hline
5 &3 &10,000 &10,000 &0.03505 &0.03303 \\ \hline
5 &5 &10,000 &10,000 &0.06631 &0.05447 \\ \hline
5 &10 &10,000 &1,000 &0.06350 &0.04341 \\ \hline
\end{tabular}
\caption{Results of counterfactual regret minimization and fictitious play in random games~\cite{Ganzfried20:Fictitious}.}
\label{ta:results-fpcfr}
\end{table*}

\section{Conclusion}
\label{se:conc}
We presented a new complete algorithm for computing Nash equilibrium in multiplayer games based on a mixed-integer quadratically-constrained feasibility program formulation. Our algorithm outperforms the previously best complete algorithms by orders of magnitude for all but the largest game class we considered. Our algorithm even has significantly smaller runtimes than the best prior incomplete methods (which also frequently fail to compute a solution). We also demonstrated that recently popular iterative algorithms have significant approximation error and are unsatisfactory for the goal of computing an exact Nash equilibrium. We ran experiments on a wide variety of game classes, and expect our algorithm to be applicable to important game models in economics, political science, security, and many other fields.

%\clearpage
\bibliographystyle{plain}
\bibliography{C://FromBackup/Research/refs/dairefs}

\end{document}